\documentclass[useAMS,usenatbib]{mn2e}
\usepackage{psfig}
\usepackage{longtable}

\def\Liso{\mbox{$L_{\rm iso}$}}
\def\Eiso{\mbox{$E_{\rm iso}$}}
\def\Ep{\mbox{$E_{\rm pk}$}}
\def\Epob{\mbox{$E_{\rm pk}^{\rm obs}$}}
\def\Ecol{\mbox{$E_{\gamma}$}}
\def\tlag{\mbox{$\tau_{\rm lag}$}}
\def\tbreak{\mbox{$t_{\rm break}$}}
\def\tbreakob{\mbox{$t_{\rm break}^{\rm obs}$}}
\def\tdur{\mbox{$T_{0.45}$}}
\def\tdurob{\mbox{$T_{0.45}^{\rm obs}$}}
\def\relE{\mbox{\Eiso--\Ep--\tbreak}}
\def\relL{\mbox{\Liso--\Ep--\tdur}}
\def\chrs{\mbox{$\chi_r^2$}}
\def\tj{\mbox{$\theta_{j}$}}

\def\Om{\mbox{$\Omega_{\rm M}$}}
\def\OL{\mbox{$\Omega_{\Lambda}$}}
\def\LCDM{\mbox{$\Lambda$CDM}}

\def\spose#1{\hbox to 0pt{#1\hss}}
\newcommand\lsim{\mathrel{\spose{\lower 3pt\hbox{$\mathchar"218$}}
     \raise 2.0pt\hbox{$\mathchar"13C$}}}
\newcommand\gsim{\mathrel{\spose{\lower 3pt\hbox{$\mathchar"218$}}
     \raise 2.0pt\hbox{$\mathchar"13E$}}}

\title[A tight correlation among prompt emission observables of long GRBs]
{Discovery of a tight correlation among the prompt emission properties
of long Gamma Ray Bursts} 

\author[Firmani et al.]{C. Firmani$^{1,2}$\thanks{E-mail:
firmani@merate.mi.astro.it},
G. Ghisellini$^{1}$, V. Avila--Reese$^{2}$ and G. Ghirlanda$^{1}$\\
$^{1}$Osservatorio Astronomico di Brera, via E.Bianchi 46, I-23807 Merate, Italy\\
$^{2}$Instituto de Astronom\'{\i}a, Universidad Nacional Aut\'onoma de M\'exico, 
A.P. 70-264, 04510, M\'exico, D.F.}

\begin{document}

% \date{Accepted 1988 December 15. Received 1988 December 14; 
% in original form 1988 October 11}

\maketitle

%\label{firstpage}

\begin{abstract}
We report the discovery of a correlation among three 
prompt emission properties of Gamma Ray Bursts.
These are the isotropic peak luminosity $L_{\rm iso}$, the 
peak energy (in $\nu L_\nu$) of the time--integrated prompt 
emission spectrum $E_{\rm pk}$, and the ``high signal" 
timescale $T_{0.45}$, previously used to
characterize the variability behavior of bursts.
In the rest frame of the source the found correlation
reads $L_{\rm iso}\propto E_{\rm pk}^{1.62}T_{0.45}^{-0.49}$.
We find other strong correlations, but at the cost
of increasing the number of variables, involving the
variability and the isotropic energy of the prompt emission.
With respect to the other tight correlations found
in Gamma Ray Bursts (i.e. between the collimation corrected
energy $E_\gamma$ and $E_{\rm pk}$, the so--called Ghirlanda
correlation, and the phenomenological correlation among
the isotropic emitted energy \Eiso, \Ep\ and the jet break
time \tbreak), the newly found correlation does not require any 
information from the afterglow phase of the bursts, 
nor any model--dependent assumption.
In the popular scenario in which   
we are receiving beamed radiation originating in a fireball pointing 
at us, the found correlation  preserves its form in the comoving frame.
This helps to explain the small scatter of the correlation, 
and underlines the role of the local brightness (i.e. the brightness
of the visible fraction of the fireball surface). 
This correlation has been found with a relatively 
small number of objects, and it is hard to establish 
if any selection bias affects it.
Its connection with the prompt local brightness
is promising, but 
a solid physical understanding is still to be found.
Despite all that, we find that some properties
of the correlation, which we discuss, support its
true existence, and this has important implications for 
the Gamma Ray Burst physics.
Furthermore, it is possible to use such correlation
as an accurate redshift estimator, and, 
more importantly, its tightness will allow 
us to use it as a tool to constrain
the cosmological parameters.
% 
% If true, this correlation can be used as an accurate redshift estimator.
% Its tightness will allow us to use it as a tool to constrain
% the cosmological parameters.
% The existence of this new correlation has implications for 
% the Gamma Ray Burst physics, which we briefly discuss.
\end{abstract}

\begin{keywords}
cosmological parameters  --- cosmology: observations --- distance 
scale --- gamma rays: bursts --- gamma rays: observations
\end{keywords}

%======================================================
\section{Introduction}
%======================================================

A major breakthrough in the understanding of Gamma Ray Burst (GRB) has
been the discovery of X--ray, optical and radio counterparts in some
long--duration bursts \citep{Costa97,vParadijs97,Frail97}.  This
allowed to measure their redshifts and realize that they are at
cosmological distances.  Since then, two potential and appealing
possibilities were cherished: the use of GRBs (i) as dust--free
tracers of the massive star formation history in the universe
\citep[e.g.,][]{Totani97,
Paczynski98,Wijers98,Lamb00,BN00,BL02,LFR02,Firmani04,Yonetoku04}, and
(ii) as cosmological standard candles able to provide a record of the
cosmic expansion history up to high redshifts
\citep[e.g.,][]{Schaefer03, GGLF04,Dai04,Firmani05,LZ05,XDL05}.  Both
items face the problem of small number of events with measured $z$,
since to determine it, deep optical/IR or X--ray spectra are needed.
Regarding the latter item (ii), the large dispersion of the GRB
energetics makes GRBs all but standard candles
\citep{Frail01,Bloom03}.  However, major advances in overcoming both
difficulties can be done on the base of tight relations that connect
GRB intrinsic energetics and/or luminosities with observed quantities.
These relations serve as redshift indicators or, when $z$ is known
independently, to make GRBs standard candles for cosmographic
purposes.

The first proposed GRB luminosity (redshift) indicators were based on
the discovery of two empirical relations between the burst
isotropic--equivalent luminosity (\Liso) and the variability ($V$),
which is a measure of the ``spikiness'' of the $\gamma$--ray light
curve \citep{FR-R00,Reichart01}, and between \Liso\ and the spectral
lag (\tlag) \citep{Norris00}.  Later, \citet{Atteia03} constructed
another redshift indicator
% proposed the search for an indicator 
based on $\gamma$--ray data alone
%  which should 
% have a negligible dependence on intrinsic energetics and a strong dependence on $z$.
% He constructed such an indicator 
linking the spectral parameters and the duration ($T_{90}$) of the
prompt emission, and he showed that it provides pseudo--redshifts
accurate within a factor of two.  In a study of the energetics of
GRBs \citet{Lloyd02a} found the evidence that GRBs with highly variable
light curves have greater $\nu F_{\nu}$ spectral peak energies. The
existence of such a correlation and of the variability--luminosity
correlation also implied that the rest frame GRB peak energy $E_{\rm
peak}$ is correlated with the intrinsic luminosity of the burst.
\citet{amati02}, by analyzing the spectra of $BeppoSAX$ GRBs, found
that the isotropic--equivalent energy radiated during the prompt phase
(\Eiso) is correlated with the rest--frame peak energy of the
$\gamma$--ray spectrum [$\Ep=\Epob(1+z)$, \Epob\ is in the observer
frame].  However, the large scatter of this correlation makes it
difficult to use it as a reliable redshift indicator.  \citet[][see
also Ghirlanda et al. 2005]{Yonetoku04} found that there is also 
some correlation between \Liso\ and $\Ep$.
% is used instead of \Eiso.

Using the above mentioned empirical correlations, hundreds of GRB
pseudo--redshifts were estimated from prompt $\gamma$--ray observables
{\it and} within a given cosmology.  
%It turns out that the
%establishment  and calibration of these correlations are cosmology
%dependent due to the lack of enough low--$z$ (calibrator) GRBs.  
However, due to the lack of enough low--$z$ (calibrator) GRBs,
these correlations are themselves cosmology dependent (but see
Ghirlanda et al. 2005a). This introduces a circularity problem if the
goal is to use these relations as distance--indicators and transform
GRBs with measured $z$ into cosmological rulers: the parameters of the
relations depend on the cosmological parameters that we pretend to
constrain.  To avoid this circular logic, \citet{Schaefer03} proposed
to determine both the GRB and cosmological parameters by fitting
simultaneously the model relations to the data, and the cosmological
model to the constructed Hubble diagram (luminosity distance $d_{\rm
L}$ vs $z$).  However, as he showed, an accurate determination of the
cosmological parameters was not possible due to the large dispersion
around the $\Liso-V$ and $\Liso-\tlag$ relations.

The hope to use GRBs as cosmological rulers renewed after the finding
by Ghirlanda, Ghisellini \& Lazzati (2004a, hereafter GGL2004) that
the Amati correlation becomes much tighter if one corrects \Eiso\ for
the collimation factor of the jet opening angle.  The presence of an
achromatic break in the GRB afterglow light curves is a strong
evidence that the GRB emission is collimated into a cone of
semiaperture angle \tj\ \citep[e.g.,][]{Rhoads97, Sari99}, where \tj\
is the rest frame time of the achromatic break in the lightcurve of
the afterglow.
% , $\tbreak=t_{\rm break}^{\rm obs}/(1+z)$, 
% if one assumes a uniform top--hat jet configuration. 
Thus, the collimated--corrected $\gamma$--ray energetic is
$\Ecol=(1-\cos\theta_j)\Eiso$. The $\Ecol$--$\Ep$ correlation,
determined with less than 20 GRBs, has been used to put constraints on
some cosmological parameters avoiding \citep{GGLF04,Firmani05} and not
avoiding \citep{Dai04} the circularity problem.

More recently, \citet[][hereafter LZ2005]{LZ05} \citep[see
also][]{Xu05} investigated the correlation among \Eiso, \Ep, and
\tbreak\ without imposing any theoretical model and assumptions.  They
found a purely empirical tight correlation among these quantities
which apparently suffers less uncertainties than the Ghirlanda
relation, though both correlations are mutually consistent (Nava et
al. 2006).  These authors used the $\Eiso$--$\Ep$--$\tbreak$
correlation as a luminosity distance indicator and applied their own
methods to avoid the circularity problem to constrain some dynamical
and kinematical cosmological parameters.

As discussed in \citet{Ghisellini05} and LZ2005, the future of GRB
cosmology is promising \citep[but see][]{FB05}, the identification of
more and more GRBs with accurately measured $z$, \Epob\ and \tbreakob\
being crucial.  Unfortunately, the determination of \tbreak\ requires
expensive follow--up campaigns involving large telescopes.  Up to now,
\tbreak\ has been measured for about one third ($\sim 20$) of the GRBs
with measured $z$ ($\sim 60$).  This motivates the need of developing
an astronomical program entirely devoted to measure $z$, \Ep\ and
\tbreak\ \citep{Lamb05}.  On the other hand, the discovery of
% but it makes also worth to search for 
tight correlations among the GRB energetics 
and the prompt $\gamma$--ray observables alone (by--passing the need
of measuring \tbreak) is of primary interest.
This is the main aim of this paper.

% In this paper, we search for prompt observables that could reduce significantly
% the dispersion of the \Liso(\Ep) or \Eiso(\Ep) correlations. 
% As a result, we find a very tight multicorrelations between \Eiso\ (\Liso) 
% and \Ep, $V$ and few more prompt quantities. 
     
In \S 2, we present the GRB sample used in this work as well as the
sample selection criteria.  Using this sample, in \S 3 we revisit the
correlation found by LZ2005 among \Eiso, \Ep\ and \tbreak.  We will
later compare this correlation with the one we find using observables
which can be extracted by the prompt emission only.  In \S 4 we
present our method to characterize the variability of the prompt
emission.  The search for a multi variable correlation among energetics
and the prompt $\gamma$--ray observables is presented and analyzed in
\S 5, where we derive the main result of our work, namely a very tight
correlation between the peak luminosity of the prompt emission \Liso,
the peak spectral energy \Ep\ and the ``high signal" timescale
$T_{0.45}$.  In \S 6 we discuss how this newly found correlation can
be used as a rather accurate redshift indicator.  The theoretical
implications of our findings are discussed in \S 7.  Finally, in \S 8
we draw our conclusions.

\section{The sample}

\begin{table*}\centering
%%\scriptsize
\begin{tabular}{@{}lllcrcrcrc@{}}
% \begin{tabular}{lllcrcrcrc}
\hline\hline
GRB    & Instrument & $z$       & Ref.& $T^{\rm obs}_{90}$ & Ref. & $T^{\rm obs}_{0.45}$ & Ref.  & $T^{\rm obs}_{\rm break}$ & Ref.\\
\hline
970228 &  SAX/WFC   & 0.695      &  1 &    80     & 27 &     2$\pm$1    & 33 &     ...          &     \\
970828 &  RXTE/ASM  & 0.957      &  2 &  146.6    & 27 &    14$\pm$2    & 33 &    2.2$\pm$0.4   & 35  \\
971214 &  SAX/WFC   & 3.42       &  3 &    35     & 27 &   7.3$\pm$0.2  &  0 &     ...          &     \\
980703 &  RXTE/ASM  & 0.966      &  4 &  102.4    & 27 &  17.2$\pm$0.3  &  0 &    3.4$\pm$0.5   & 35  \\
990123 &  SAX/WFC   &  1.6       &  5 &   100     & 27 &  17.2$\pm$0.1  &  0 &    2.0$\pm$0.5   & 36  \\
990506 &  BAT/PCA   & 1.307      &  6 &   220     & 27 & 14.05$\pm$0.06 &  0 &     ...          &     \\
990510 &  SAX/WFC   & 1.619      &  7 &    75     & 27 &  5.00$\pm$0.05 &  0 &    1.6$\pm$0.2   & 37  \\
990705 &  SAX/WFC   & 0.843      &  8 &    42     & 27 &     5$\pm$2    & 33 &    1.0$\pm$0.2   & 35  \\
990712 &  SAX/WFC   & 0.43       &  7 &    20     & 27 &     ...        & 33 &    1.6$\pm$0.2   & 38  \\
991216 &  BAT/PCA   & 1.02       &  9 &    24.9   & 27 &  3.78$\pm$0.02 &  0 &    1.2$\pm$0.4   & 35  \\
000131 & Uly/KO/NE  & 4.5        & 10 &  110.1    & 27 &     4$\pm$1    & 33 &     ...          &     \\
000911 & Uly/KO/NE  & 1.058      & 11 &  500      & 27 &   5.2$\pm$0.5  & 34 &     ...          &     \\
011211 &  SAX/WFC   & 2.14       & 12 &  ...      &    &     ...        &  0 &    1.5$\pm$0.2   & 39  \\
020124 &    HETE    & 3.2        & 13 &    50     & 28 &  14.0$\pm$1.5  &  0 &    3.0$\pm$0.4   & 40  \\
020813 &    HETE    & 1.25       & 14 &    90     & 27 &    17$\pm$1    &  0 &   0.43$\pm$0.06  & 41  \\
021211 &    HETE    & 1.01       & 15 &    13     & 28 &  0.81$\pm$0.05 &  0 &     ...          &     \\
030226 &    HETE    & 1.98       & 16 &   138     & 28 &    26$\pm$2    &  0 &   0.84$\pm$0.10  & 41  \\
030328 &    HETE    & 1.52       & 17 &   316     & 28 &  26.2$\pm$0.7  &  0 &    0.8$\pm$0.1   & 42  \\
030329 &    HETE    & 0.1685     & 18 &    33     & 28 &   5.0$\pm$0.1  &  0 &    0.5$\pm$0.1   & 43  \\
030429 &    HETE    & 2.658      & 19 &    77     & 28 &     ...        &    &    1.8$\pm$1.0   & 44  \\
040924 &    HETE    & 0.859      & 20 &     5     & 21 &  0.45$\pm$0.02 &  0 &     ...          &     \\
041006 &    HETE    & 0.716      & 21 &    25     & 21 &  4.47$\pm$0.01 &  0 &   0.16$\pm$0.04  & 45  \\
050525 &   SWIFT    & 0.606      & 22 &     8.8   & 29 &  2.09$\pm$0.01 &  0 &   0.15$\pm$0.01  & 30  \\
\hline
980425 & BAT/SAX    & 0.0085     & 23 &    37.4   & 30 &   7.5$\pm$0.2  &  0 &     ...          &     \\
990712 &    SAX     & 0.433      & 24 &     20.0  & 31 &     4$\pm$1    &  0 &    1.6$\pm$0.2   &  46 \\
010921 &    HETE    & 0.45       & 25 &   24.6    & 21 &   7.3$\pm$0.4  &  0 &     ...          &     \\
031203 &  INTEGRAL  & 0.106      & 26 &   30      & 32  &     5$\pm$1   &  0 &     ...          &     \\
% GRB    instrument     z         ref     T90      ref        T45        ref         Tbrk         ref
\hline\hline
\end{tabular}
\caption{$T^{\rm obs}_{90}$, $T^{\rm obs}_{0.45}$, $T^{\rm obs}_{\rm break}$ are expressed in seconds.
References: 
  (0) This paper;
  (1) Djorgovski et al. 1997;
  (2) Djorgovski et al. 2001;
  (3) Kulkarni et al. 1998;
  (4) Djorgovski et al. 1998; 
  (5) Hjorth et al. 1999; 
  (6) Bloom et al. 2001;
  (7) Vreeswijk et al. 2001; 
  (8) Amati et al. 2000; 
  (9) Vreeswijk et al. 1999; 
  (10) Andersen et al. 2000;
  (11) Price et al. 2002;
  (12) Amati et al. 2004; 
  (13) Hjorth et al. 2003; 
  (14) Barth et al. 2003; 
  (15) Vreeswijk et al. 2003;
  (16) Greiner et al. 2003;
  (17) Rol et al. 2003; 
  (18) Greiner et al. 2003a; 
  (19) Weidinger et al. 2003; 
  (20)  Wiersema et al. 2004;
  (21) http://space.mit.edu.HETE/Bursts; 
  (22) Foley et al. 2005; 
  (23)  Tinney et al. 1998;
  (24)  Galama et al. 1999;
  (25)  Djorgovski et al. 2001a;
  (26)  Bersier et al. 2003;
  (27) Ghirlanda et al. 2004a; 
  (28) Sakamoto et al. 2005;  
  (29) Blustin et al. 2005;
  (30) Jimenez et al. 2001;
  (31) Amati et al. 2002;
  (32) Mereghetti \& Gotz 2003;
  (33) Reichart et al. 2001; 
  (34) Guidorzi et al. 2005;
  (35) Bloom et al. 2003; 
  (36) Kulkarni et al. 1999; 
  (37) Israel et al. 1999; 
  (38) Bjornsson et al. 2001;
  (39) Jakobsson et al. 2003; 
  (40) Berger et al. 2002; 
  (41) Klose et al. 2004; 
  (42) Andersen et al. 2003; 
  (43) Berger et al. 2003; 
  (44) Jakobsson et al. 2004; 
  (45) Stanek et al. 2005;
  (46) Bjornsson et al. 2001.}
\end{table*}

The study of the correlations among prompt emission observables requires 
a careful selection of the GRB sample in order to work with a reliable 
and homogeneous set of information. 
The basic selection criteria we adopt are the knowledge of: 
\begin{enumerate}
\item
the redshift $z$;
\item the peak flux $P$ and the fluence $F$, better if defined both in
the same energy range;
\item the $\nu F_{\nu}$ peak energy, \Epob, and a fitting to the spectrum 
able to provide a reasonable determination of the bolometric correction.
Note that \Epob\ is derived in all cases using the time integrated spectrum.
\item \tdurob\ and $V$. When possible, these parameters have been calculated 
with our own code using the published light curves (see below). 
\end{enumerate}
The time \tdurob\ is the prompt ``high--signal'' timescale, introduced by
\citet[][hereafter R2001]{Reichart01}, defined in the rest energy
range 50--300 keV (see \S 4.1).  For comparative purposes, the
knowledge of the observed \tbreakob\ is also desirable, although we
did not require it as a selection criterion.  For the few cases when
the observed $P$ is not available, we have estimated it from the
light curve profile by using the published spectrum and fluence.

Our sample includes mainly the GRBs reported in GGL2004 (15 GRBs); in
some cases the data of these bursts have been updated with more
recent published analysis.  To the GGL2004 sample, we have added 7
GRBs.  In Table 1 we present the resulting sample of the 22 GRBs used
in this paper, specifying the mission/instrument  from which
the data were acquired and the spectroscopically measured redshift $z$
.  For each burst we report the duration $T_{90}$ , the
$T^{\rm obs}_{0.45}$ timescale  and the jet break time $T^{\rm
obs}_{\rm break}$  in the observer frame. In most cases
$T^{\rm obs}_{0.45}$ was computed from the GRB light curve as
discussed below (\S 4.1). The data from R2001 where adapted to the
50-300 keV band by interpolation. We have compared our results with the ones
of R2001 for the common GRBs finding a good agreement. This comparison
has been helpful for us to estimate the uncertainty of the data derived 
from R2001.

In Table 2, we list the bolometric corrected fluence ($F$) and
peak flux ($P$) and the corresponding energy range where they were
computed.  In Table 2 we also report the spectral parameters
$\alpha$ and $\beta$ for those bursts whose spectrum was fitted by the
\citet{Band93} spectral model and only the $\alpha$ spectral parameter
when the spectrum was fitted with a cutoff--powerlaw model. The
fluence and the peak flux are given respectively either in units of 
erg cm$^{-2}$ and erg cm$^{-2}$ s$^{-1}$, in round brackets, or in
units of photon cm$^{-2}$ and photons cm$^{-2}$ s$^{-1}$, in square
brackets. This choice of homogeneity, as well as the one of the same energy 
range for the fluence and the peak flux, have been made with the aim to 
reduce the effects of the spectral uncertainty on the $\Liso/\Eiso$ ratio. 
%
%
%In other words, we take care to use the originally reported units of
%the data on both the peak flux and fluence.  The use of the same units for the
%fluence and peak flux, in fact, allows us to minimize the error on the
%ratio \Liso/\Eiso.
%

In Table 3 we report the rest frame $\Ep$, $\Eiso$, $\Liso/\Eiso$
ratio and $V$. 
The sample has been divided into two classes.  Class 1 (16 events)
comprises those GRBs with a good determination of the bolometric
correction.  Class 2 (5 events) comprises those events with uncertain
bolometric corrections.  In fact, the uncertain high-energy spectrum of Class 2
bursts does not allow to constrain the high--energy Band spectral
index ($\beta$).  In these cases spectral fits were performed either
freezing $\beta$ to a fiducial value ($\beta=-2.3$) or the fit was
performed using a cutoff power law model.  For GRB~050525 we assign a
special class, 3, because the determination of $\beta$ in this case is
highly uncertain.

Regarding the uncertainties, for those quantities taken from the
observations ($z$, $F$, $P$, and \Epob), we use the errors reported in
the literature.  For the derived quantities \Eiso, \Liso, and $\Phi$
(see below), we calculate their uncertainties by propagating the
errors under the assumption that they are uncorrelated.

The conventional \LCDM\ cosmology with $\Om=0.3$, $\OL=0.7$ and $h=0.7$ 
is assumed here for the calculation of luminosity distances.

% ------------------------------------------------------

\begin{figure}
\psfig{file=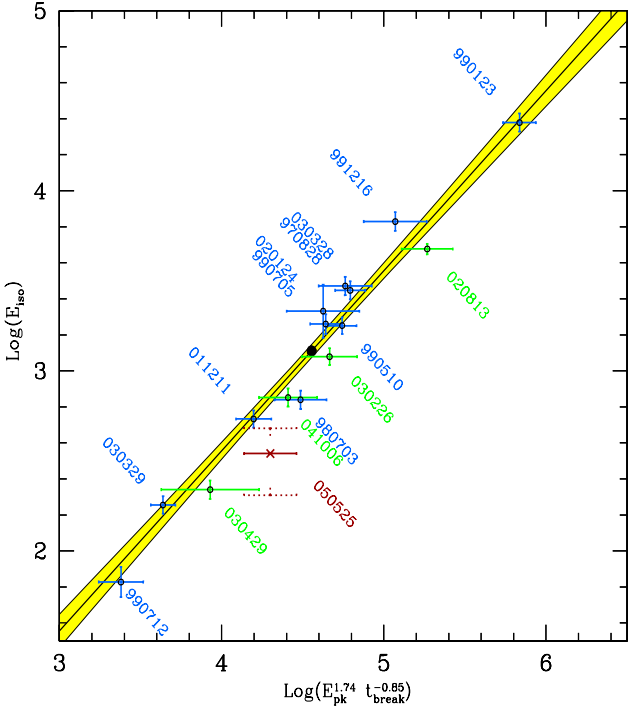,height=10cm,width=9cm}
\caption{Isotropic emitted energy $E_{\rm iso}$ as a function of
$E_{\rm peak}^{1.74} t_{\rm break}^{-0.85}$ (Eq. 1.).  The open
circles represent the 15 bursts (i.e. all those with measured
$t_{\rm break}$ in Tab.~1 except 050525) used for the fit (solid
line). The solid filled region represents the 1$\sigma$ uncertainty on
the best fit (see Eq.~1). The fit is performed in the centroid defined
by the data points (solid circle) where the errors on the best fit
parameters are uncorrelated.
%
% Multiple variable linear regression analysis for the sample of 15 GRBs
% labeled by empty circles. 
%The best fit is obtained  fixing the centroid (solid circle)
%defined by the two coordinate errors in order to garantee uncorrelated errors
%for the fitted parameters. 
The color codes are: blue for class 1 bursts, green for class 2 and 
brown for class 3.  For GRB~050525 we also report 
the values obtained assuming $\beta=-2.1$ and $\beta=-5$ (dotted lines).
\label{fig1}}
\end{figure}

%%%%%%%%%%%%%%%%%%%%%%%%%%%%%%%%%%%%%%%%%%%%%%%%%%%%%%%%%%%%%%%%%%%%
\begin{table*}\centering
\begin{tabular}{@{}lrrrrcrrc@{}}
%%\scriptsize
\hline\hline
GRB    & $\alpha$       &  $\beta$       & fluence           & range   & ref & peak flux       & range     & ref\\
\hline
% GRB          alfa            beta            fluence            range   ref     peak flux           range    ref
970228 & -1.54$\pm$0.08 &  -2.5$\pm$0.4  & (1.1$\pm$0.1)e-5   &  40-700 &  1  & (3.7$\pm$0.8)e-6  &   40-700  &  4 \\
970828 & -0.70$\pm$0.08 &  -2.1$\pm$0.4  & (9.6$\pm$0.9)e-5   & 20-2000 &  2  & (5.9$\pm$0.3)e-6  & 30-10$^4$ & 10 \\
971214 & -0.76$\pm$0.10 &  -2.7$\pm$1.1  & (8.8$\pm$0.9)e-6   &  40-700 &  1  & (6.8$\pm$0.7)e-7  &   40-700  &  4 \\
980703 & -1.31$\pm$0.14 & -2.39$\pm$0.26 & (2.3$\pm$0.2)e-5   & 20-2000 &  2  & (1.6$\pm$0.2)e-6  &   50-300  &  2 \\
990123 & -0.89$\pm$0.08 & -2.45$\pm$0.97 & (3.0$\pm$0.4)e-4   &  40-700 &  3  & (1.7$\pm$0.5)e-5  &   40-700  &  4 \\
990506 & -1.37$\pm$0.15 & -2.15$\pm$0.38 & (1.9$\pm$0.2)e-4   & 20-2000 &  1  &   [18.6$\pm$0.1]  &   50-300  & 11 \\
990510 & -1.23$\pm$0.05 &  -2.7$\pm$0.4  & (1.9$\pm$0.2)e-5   &  40-700 &  3  & (2.5$\pm$0.2)e-6  &   40-700  &  4 \\
990705 & -1.05$\pm$0.21 &  -2.2$\pm$0.1  & (7.5$\pm$0.8)e-5   &  40-700 &  3  & (3.7$\pm$0.1)e-6  &   40-700  &  4 \\
990712 & -1.88$\pm$0.07 & -2.48$\pm$0.56 & (6.5$\pm$0.3)e-6   &  40-700 &  3  &   [4.1$\pm$0.3]   &   40-700  & 12 \\
991216 & -1.23$\pm$0.13 & -2.18$\pm$0.39 & (1.9$\pm$0.2)e-4   & 20-2000 &  2  &  [67.5$\pm$0.2]   &   50-300  & 11 \\
000131 & -0.69$\pm$0.08 & -2.07$\pm$0.37 & (4.2$\pm$0.4)e-5   & 20-2000 &  1  &  [7.89$\pm$0.08]  &   50-300  & 13 \\
000911 & -1.11$\pm$0.12 &  -2.3$\pm$0.4  & (2.2$\pm$0.2)e-4   & 15-8000 &  1  & (2.0$\pm$0.2)e-5  &  15-8000  & 14 \\
011211 & -0.84$\pm$0.09 &                &     ...            & ...     &  4  & (5.0$\pm$1.0)e-8  &   40-700  & 15 \\
020124 & -0.87$\pm$0.17 &  -2.7$\pm$0.5  &    [166$\pm$13]    &  2-400  & 5,6 &  [9.4$\pm$1.8]    &    2-400  &  5 \\
020813 &  -0.9$\pm$0.1  &                &   [1325$\pm$23]    &   2-400 &  5  &  [32.3$\pm$2.1]   &    2-400  &  5 \\
021211 & -0.86$\pm$0.09 & -2.23$\pm$0.20 &     [93$\pm$3]     &   2-400 &  5  &    [30$\pm$2]     &    2-400  &  5 \\
030226 & -0.83$\pm$0.16 &      -2.3      &    [114$\pm$11]    &   2-400 &  5  &   [2.7$\pm$0.6]   &    2-400  &  5 \\
030328 & -1.14$\pm$0.03 &  -2.2$\pm$0.3  &    [751$\pm$12]    &   2-400 &  5  &  [11.6$\pm$0.9]   &    2-400  &  5 \\
030329 & -1.26$\pm$0.02 & -2.28$\pm$0.06 &   [4963$\pm$42]    &   2-400 &  5  &    [451$\pm$25]   &    2-400  &  5 \\
030429 & -1.12$\pm$0.24 &                &     [40$\pm$5]     &   2-400 &  5  &   [3.8$\pm$0.8]   &    2-400  &  5 \\
040924 & -1.17$\pm$0.05 &                & (2.7$\pm$0.1)e-6   &  20-500 & 7,8 & (2.6$\pm$0.3)e-6  &   20-500  &  0 \\
041006 &  -1.4$\pm$0.1  &                & (2.0$\pm$0.2)e-5   &  25-100 &  7  & (1.0$\pm$0.1)e-6  &   25-100  &  0 \\
050525 &  0.00$\pm$0.12 &  -2.3$\pm$0.1  & (2.01$\pm$0.05)e-5 &  15-350 &  9  &   [47.7$\pm$1.2]  &   15-350  &  9 \\
\hline
980425 &  -1.0$\pm$0.3  &  -2.1$\pm$0.1  & (3.8$\pm$0.4)e-6   & 20-2000 & 16  &    [0.4$\pm$0.1]  &   50-300  & 17 \\
990712 & -1.88$\pm$0.07 & -2.48$\pm$0.56 & (6.5$\pm$0.3)e-6   &  40-700 &  1  &    [4.1$\pm$0.3]  &   40-700  &  1 \\
010921 & -1.49$\pm$0.16 &      -2.3      & (1.8$\pm$0.1)e-5   &   2-400 &  5  &     [40$\pm$4]    &    2-400  &  5 \\
031203 &  -1.63$\pm$0.1 &                & (2.0$\pm$0.4)e-6   &  20-200 & 18  & (2.4$\pm$0.2)e-7  &   20-200  & 18 \\
\hline\hline
% GRB          alfa            beta            fluence            range            peak flux          range    ref
\end{tabular}
\caption{$\alpha$ and $\beta$ represent the photon spectral index of
the model spectrum. The fluence is given in units of erg cm$^{-2}$
(round brackets) or in units of photon cm$^{-2}$ (square brackets). The
peak flux is given in units of erg cm$^{-2}$ s$^{-1}$ (round brackets)
or photon cm$^{-2}$ s$^{-1}$ (square brackets). The energy range of the
fluence and peak flux is expressed in units of keV.  
References: 
(0) This paper; 
(1) Ghirlanda et al. 2004a and references therein; 
(2) Jimenez et al. 2001; 
(3) Amati et al. 2002; 
(4) Amati et al. 2004; 
(5) Sakamoto et al. 2005; 
(6) Atteia et al. 2005; 
(7) http://space.mit.edu.HETE/Bursts; 
(8) Golenetskii et al. 2004; 
(9) Blustin et al. 2005; 
(10) Yonetoku et al. 2004; 
(11) The BATSE catalogue (Paciesas et al. 1999); 
(12) Frontera et al. 2001;
(13) Andersen et al. 2000; 
(14) Price et al. 2002; 
(15) Piro et al. 2005;
(16) Yamazaki et al. 2003; 
(17) Paciesas et al. 1999;
(18) Sazonov et al. 2004.}
\end{table*}

\begin{table*}\centering
\begin{tabular}{@{}lcrcrcrrc@{}}
%%\scriptsize
\hline\hline
GRB    & class & E$_{pk}$  & ref  & E$_{iso}$ & ref  & L$_{iso}$/E$_{iso}$   & Var & ref\\
\hline
% GRB  class      Epk      ref         Eiso          ref         L/E                 Var           ref
970228 & 1 &  195$\pm$64  &  1 & (1.60$\pm$0.12)e52 &  1 &  (5.7$\pm$1.3)e-1   &  0.08$\pm$0.05  &  8  \\
970828 & 1 &  583$\pm$116 &  2 & (2.96$\pm$0.35)e53 &  1 &  (8.5$\pm$2.4)e-2  &  0.10$\pm$0.01   &  8  \\
971214 & 1 &  685$\pm$133 &  1 & (2.11$\pm$0.24)e53 &  1 & (34.2$\pm$5.0)e-2  & 0.087$\pm$0.004  &  0  \\
980703 & 1 &  502$\pm$100 &  2 &  (6.9$\pm$0.8)e52  &  1 & (30.3$\pm$6.1)e-2  & 0.064$\pm$0.003  &  0  \\
990123 & 1 & 2030$\pm$160 &  3 &  (2.4$\pm$0.3)e54  &  1 & (14.7$\pm$4.8)e-2  & 0.129$\pm$0.001  &  0  \\
990506 & 1 &  653$\pm$130 &  1 &  (9.5$\pm$1.1)e53  &  1 &  (4.4$\pm$1.3)e-2  & 0.326$\pm$0.001  &  0  \\
990510 & 1 &  423$\pm$42  &  3 &  (1.8$\pm$0.2)e53  &  1 & (34.0$\pm$4.6)e-2  & 0.229$\pm$0.002  &  0  \\
990705 & 1 &  348$\pm$28  &  3 & (1.82$\pm$0.23)e53 &  1 &  (9.1$\pm$1.0)e-2  &   0.15$\pm$0.05  &  8  \\
990712 & 1 &   93$\pm$15  &  3 &  (6.7$\pm$1.3)e51  &  1 & (11.1$\pm$1.9)e-2  &      ...         &     \\
991216 & 1 &  641$\pm$128 &  2 & (6.75$\pm$0.81)e53 &  1 & (16.8$\pm$5.2)e-2  & 0.124$\pm$0.001  &  0  \\
000131 & 1 &  714$\pm$142 &  1 & (1.84$\pm$0.22)e54 &  1 &  (7.7$\pm$2.9)e-2  &   0.11$\pm$0.01  &  8  \\
000911 & 1 & 1190$\pm$238 &  1 &  (8.8$\pm$1.0)e53  &  1 & (18.7$\pm$2.5)e-2  & 0.077$\pm$0.034  &  9  \\
011211 &   &  186$\pm$24  &  4 &  (5.4$\pm$0.6)e52  &  1 &      ...           &     ...          &     \\
020124 & 1 &  390$\pm$113 &  5 & (2.15$\pm$0.73)e53 &  0 & (23.8$\pm$4.9)e-2  &  0.29$\pm$0.04   &  0  \\
020813 & 2 &  478$\pm$95  &  6 &  (4.7$\pm$0.3)e53  &  0 & (54.8$\pm$3.7)e-3  &  0.247$\pm$0.004 &  0  \\
021211 & 1 &   92$\pm$14  &  7 & (1.10$\pm$0.13)e52 &  1 & (64.8$\pm$4.8)e-2  &  0.006$\pm$0.004 &  0  \\
030226 & 2 &  289$\pm$63  &  7 & (1.20$\pm$0.13)e53 &  1 &  (7.1$\pm$1.7)e-2  &  0.08$\pm$0.04   &  0  \\
030328 & 1 &  318$\pm$34  &  7 & (2.80$\pm$0.33)e53 &  1 & (38.9$\pm$3.1)e-3  &  0.052$\pm$0.006 &  0  \\
030329 & 1 &   79$\pm$3   &  7 &  (1.8$\pm$0.2)e52  &  1 & (10.6$\pm$0.6)e-2  &  0.104$\pm$0.002 &  0  \\
030429 & 2 &  128$\pm$26  &  7 & (2.19$\pm$0.26)e52 &  1 & (34.7$\pm$8.5)e-2  &      ...         &     \\
040924 & 2 &  125$\pm$12  & 10 &  (7.3$\pm$0.8)e51  &  0 & (18.6$\pm$2.9)e-1  &  0.007$\pm$0.001 &  0  \\
041006 & 2 &  109$\pm$22  & 11 &  (7.1$\pm$0.8)e52  &  0 &  (8.6$\pm$1.2)e-2  &  0.082$\pm$0.003 &  0  \\
050525 & 3 &  127$\pm$6   & 12 &  (3.5$\pm$0.5)e52  &  0 & (24.7$\pm$1.6)e-2  &  0.099$\pm$0.001 &  0  \\
                                \hline
980425 &   &  119$\pm$24  &  1 &  (1.6$\pm$0.2)e48  &  1 & (4.18$\pm$1.19)e-2 & 0.0055$\pm$0.005 &  0  \\
990712 &   &   93$\pm$16  &  1 & (6.72$\pm$1.29)e51 &  1 & (1.11$\pm$0.19)e-1 &  0.042$\pm$0.017 &  0  \\
010921 &   &  154$\pm$30  &  1 &  (1.5$\pm$0.1)e52  &  0 & (8.30$\pm$1.00)e-2 &  0.010$\pm$0.009 &  0  \\
031203 &   &    $>$210    & 13 &  (1.0$\pm$0.4)e50  &  0 & (1.33$\pm$0.30)e-1 &    0.1$\pm$0.01  &  0  \\
                             \hline\hline
% GRB   type      Epk       ref         Eiso          ref          L/E                  Var        ref
\end{tabular}
\caption{The peak energy $E_{\rm pk}$ and the bolometric isotropic
equivalent energy $E_{\rm iso}$ are expressed in units of keV and erg,
respectively. $E_{\rm iso}$ and $L_{\rm iso}$ are calculated in the
1--10$^4$ keV rest frame energy band. The Variability (Var) is
calculated as described in the text (\S 4). 
References: 
(0) This paper
(1) Ghirlanda et al 2004a (references therein); 
(2) Jimenez et al. 2001;
(3) Amati et al. 2002; 
(4) Amati et al. 2004;
(5) Atteia et al. 2005;
(6) Barraud et al. 2003;
(7) Sakamoto et al. 2005; 
(8) Reichart et al. 2001;
(9) Guidorzi et al. 2005;
(10) Golenetskii et al. 2004;
(11) http://space.mit.edu.HETE/Bursts;
(12) Blustin et al. 2005;
(13) Sazonov et al. 2004;
}

\end{table*}

% =======================================================
\section{The \relE\ correlation}
%\section{The $E_{\rm iso}$--$E_{\rm pk}$--$t_{\rm break}$ correlation}
% =======================================================

As stated in the Introduction, our aim is to find a statistically
significant (multi)variable correlation among GRB prompt emission
observables.  Previous to this, we present here for our sample the
% tightest GRB 
% multivariable correlation reported in the literature, the one between 
\Eiso, \Ep\ and \tbreak\ correlation (LZ2005), and discuss some of its
implications.  We will also attempt later to compare and connect our
results to this correlation.  Using the same fitting procedure as in
Nava et al. (2006)
% A multiple variable linear regression analysis on our sample 
for those 15 events for which \tbreak\ is known, we obtain the
following best fit, in the GRB rest frame:
% with errors on two coordinates gives in the GRB rest frame 
% the following best fit:
%
\begin{eqnarray} 
\Eiso \, &=& \, 10^{53.11\pm0.04}
\left({\Ep \over {10^{2.48} {\rm keV}}}\right)^{1.74\pm0.10}
\nonumber \\
&~&
\left({\tbreak \over {10^{-0.29} {\rm d}}}\right)^{-0.85\pm0.15} {\rm erg}, 
\label{Liang} 
\end{eqnarray} 
%EQUAZIONE VERIFICATA
with \chrs=0.67 (for 12 dof).  Notice that, differently from LZ2005,
we do take into account for the fit the uncertainties in the data.
The data and Eq. (\ref{Liang}) are shown in a two dimensional plot in
Fig. \ref{fig1}.  In our sample, which has been constructed to find
correlations among prompt variables alone, GRB~020405 and GRB~021004
are not included because for them the high--energy spectral parameter
$\beta$ is larger than $-2$ (GRB~020405) or unknown (GRB~021004),
\citep[but see][]{nava06}.  We did not take into account GRB~050525
for the fit of Eq. (\ref{Liang}) due the large uncertainty in the
$\beta$ parameter, but we plot it in Fig. \ref{fig1}: the
dotted lines correspond to values of $\beta$ between $-2.1$ and
$-5.0$.

The multi variable correlation given in Eq. (\ref{Liang}) not only
establishes a link between the $\gamma$--ray prompt and the afterglow,
but it may carry valuable information on the jet opening angle
$\theta_j$.  In the context of the fireball uniform jet model, this
angle (supposed to be the same in the prompt and in the afterglow) is
related to \tbreak, the achromatic break time of the afterglow light
curve \citep{Rhoads97,Sari99}. Depending on the circumstellar medium
distribution, two models have been proposed.  For the homogeneous
medium (HM) with number density $n$,
\begin{equation} 
\theta_j = 0.161 \left({\tbreak}\right)^{3/8}
\left({{n \eta_\gamma} \over \Eiso}\right)^{1/8} \ \ {\rm (HM)}. 
\label{hm} 
\end{equation} 
For a wind medium (WM) with a density profile 
$n(r) = 5 \times 10^{11} A r^{-2}$ g cm$^{-2}$
(here $A=1$ corresponds to the mass loss rate due to a wind 
of $\dot M_{\rm w} =10^{-5}
M_\odot$ yr$^{-1}$ and a wind velocity $v_{\rm w}=10^3$ km s$^{-1}$),
we have
\begin{equation}
\theta_j = 0.202 \left({\tbreak}\right)^{1/4}
\left({{A \eta_\gamma} \over \Eiso}\right)^{1/4} \ \ {\rm (WM)},
\label{wm} 
\end{equation} 
where $\theta_j$ is expressed in radians and $\eta_\gamma$ is the
efficiency of the emission, i.e. the fraction of the kinetic energy of
the fireball which is emitted in the prompt $\gamma$--ray phase.  Such
estimate of $\theta_j$ implies the knowledge of $z$ and of the
luminosity distance.

Interestingly enough, by introducing the found multi variable
correlation Eq. (\ref{Liang}) in Eqs. (\ref{hm}) and (\ref{wm}), one
may express \tj\ in terms of prompt observables.  Thus, we obtain
approximately that:
\begin{equation} 
\theta_j^2 \propto {\Ep^{3/2} \over \Eiso} \ \ {\rm (HM)};
\ \ \ \ \ \theta_j^2 \propto {\Ep \over \Eiso} \ \ {\rm (WM)}. 
\label{hw1} 
\end{equation} 
Defining the collimation corrected energy approximately as
\Ecol=\Eiso$\theta_j^2$, Eqs. (\ref{hw1}) lead to
\begin{equation} 
\Ecol \propto \Ep^{3/2} \ \ {\rm (HM)};
\ \ \ \ \ \Ecol \propto \Ep \ \ {\rm (WM)}. 
\label{hw2} 
\end{equation}
These relations were extensively discussed by GGL2004 and
\citet{nava06}.  Finally, if we identify \Ecol/\Ep\ with the total
number of photons $N_\gamma$ emitted by the prompt, then an
interesting conclusion follows \citep{nava06} :
\begin{equation} 
N_\gamma \propto \Ep^{1/2} \ \ {\rm (HM)};
\ \ \ \ \ N_\gamma = const \sim 10^{57} \ \ {\rm (WM)}. 
\label{hw3} 
\end{equation} 
In \S 7 we will use these results to compare the \relE\ or
Ghirlanda correlation with the new one we will introduce in \S 5.

%========================================================
\section{Characterizing the variability of the prompt emission}
%========================================================

The starting point, as mentioned in the Introduction, are the
\Liso--\Ep\ or the \Eiso--\Ep\ correlations.  Our aim is to see if the
introduction of more prompt variables can produce significantly
tighter correlations than these ones.  Among the most relevant
information from the $\gamma$--ray prompt emission is the lightcurve.

The variability $V$, a quantity that estimates the ``spikiness'' of
the lightcurve \citep[see for details][]{FR-R00}, offers a way to
partially quantify the information contained in the light curves.

Here we introduce another variability indicator making use of the
fluence $F$ to the peak flux $P$ ratio.  Such ratio gives a time,
which compared with a burst timescale provides a further
characterization of the lightcurve pattern variability.  We calculate
the $F/P$ ratio using the {\it bolometric corrected} quantities (in the 
range of $1-10^4$ keV).  Using the prompt ``high--signal'' timescale, 
\tdurob\ (see next subsection) we define, in the GRB rest frame  
\begin{equation} 
\Phi = \frac{\Liso  \tdur}{ \Eiso},
\label{fi} 
\end{equation} 
where $\tdur=\tdurob/(1+z)$. $\Phi$ is a scalar with a clear meaning
in the observer frame, the GRB rest frame, and the fireball comoving
frame.  If one multiplies the numerator and denominator of the right
term by $\theta_j^2/\Ep$, then $\Phi$ reveals itself as the ratio of
the number of photons emitted during the ``high-signal'' regime
to the total number of emitted photons, which is particularly 
interesting in the comoving frame. A further advantage is that $\Phi$ may 
be derived from observables easy to handle.

\subsection{Definition of variability}

The definition of the variability 
$V$ requires the estimate of a smoothing time scale,
\tdurob, that, following R2001, we assume to be {\it the time spanned
by the brightest 45$\%$ of the total counts above the background}. In
practice, we estimate from the light curve the fraction of counts and the
total time when
the signal exceeds a given threshold.  We identify the threshold when
this fraction is 0.45, and we calculate the total time during which
the signal exceeds this threshold.  The assumed energy range is
$50-300$ keV at rest. We use the recipe proposed by R2001 to pass from
the observed energy range to the assumed rest one.  Finally, we use
the lightcurve time binning of {\it HETE-II}, 164--ms.  Though for
most of the $BATSE$ GRBs the time bin is 64--ms, we preferred to take
the larger {\it HETE-II} time bin for uniformity.  In Table 1 we
report the \tdurob\ calculated by us from the publicly available
lightcurves.  In the few cases when the lightcurves were not
available, we have used the time scales \tdurob\ reported in R2001
and, for one event (see below), in \citet{Guidorzi05}.

The $V$ parameter decreases systematically with the S/N ratio.  In
order to handle a quantity characterizing the signal it is necessary
to disentangle the noise contribution.  To this purpose R2001
suggested an analytic approximation.  We introduce here an alternative
method, with the aim to reduce the uncertainty on the variability
noise correction avoiding any approximation.  Our method is especially
useful for data with a low S/N ratio.  Based on the assumption that
the photon shot noise obeys a Poissonian distribution, we use a Monte 
Carlo approach to simulate the noise contribution to the variability.

The basics of our approach is as follows.  Given a noiseless signal
({\it zero order}) with an {\it intrinsic variability}, we can
introduce a {\it first order} realization of the signal by distorting
it with its own Poisson noise. The {\it first order variability} will
be different from the {\it intrinsic variability}. Then we construct a
{\it second order} realization of the signal distorting the previous
{\it first order} realization with its own Poisson noise. This {\it
second order} realization will have a new variability.  {\it Higher
order} realizations may be carried out up to a fixed number.  The
sequence of realizations may be randomly repeated and the average
behavior and scatter of the variability sequences may be obtained. A
simple statistical analysis allows to recover the {\it intrinsic
variability} with some uncertainty from the higher (first, second,...)
order realizations. Numerical experiments up to the third order
realizations have been carried out successfully on artificial signals.
We have found that the simple linear extrapolation $V_0=2 V_1-V_2$ 
is enough to obtain a reasonable zero order estimate (here the 
index identifies the order).
In the case of a real signal, the {\it zero order} signal is unknown
while its {\it intrinsic variability} is the result we are looking
for. Actually we have just a {\it first order} realization with its
intrinsic Poisson noise.  Starting from here we apply the previous
scheme building several times the {\it second order} realizations. The
final result is the {\it intrinsic variability} with its uncertainty.

We have taken some care to test our method using a real GRB light curve.
To this aim, we have assumed that the observed light curve is 
a pure zero order signal (i.e. ignor any Poisson noise
and call $V_0$ the variability of this light curve). 
Then we have distorted this signal once with a Poisson noise obtaining 
a first order realisation ($\tilde V_1$), and finally on this one 
we have applied our method 
(i.e. derive $\tilde V_2$ by a Monte Carlo method, extrapolate back
and estimate the average $\tilde V_0$ and its error). 
The result may be compared now with the
variability of the ``zero" order signal $V_0$. 
In the case of the rather noisy 
GRB 971214 light curve, $V_0=0.1169$ lies inside
the obtained $\tilde V_0=0.116 \pm 0.004$ (1$\sigma$ error), 
while for the high signal GRB 991216 light curve we 
have $V_0=0.1438$, which is consistent with $\tilde V_0=0.1445\pm0.0010$.
  
In principle our variability estimate follows a different procedure
from the original method proposed by R2001, therefore a comparison
between the two methods is desirable.  We have used the GRBs of our
sample in common with the R2001 sample to compare the values of \tdurob\
and $V$. From the results given in Table 1 and Table 3 it can be seen
that our method gives \tdurob\ and $V$ in general good agreement with
those published by R2001.  The only difference is a smaller
uncertainty in our estimate of $V$.
%More importantly, our method allows to calculate the uncertainty on \tdurob. 
% this fact allowed us to associate a rough uncertainty 
% to the \tdurob\ estimated by R2001. 
The fact that our results agree with R2001 justifies the inclusion in
our sample of four GRBs whose $V$ and \tdurob\ have been estimated by
R2001 only.
% to our sample without the risk to brake the homogeneity in the sample. 

In a recent paper, \citet{Guidorzi05} have presented new calculations
of $V$ for a sample of 32 GRBs.  For some GRBs in common with our
sample, the processed lightcurves used by these authors could differ
from the ones used by us.  Besides, for the $Beppo$SAX/WFC events they
apply two further ``instrumental'' corrections in the calculation of
$V$ that we did not apply because of the lack of information.
According to the authors, these corrections affect $V$ significantly
only for relatively short GRBs exhibiting sharp intense pulses.  On
the other hand, our calculation of $V$ and its uncertainty includes a
new method to treat the noise correction that improves the results
especially for data with low S/N ratios, while \citet{Guidorzi05} use
the same analytical approach of R2001.  We have compared \tdurob\ and
$V$ for the 5 GRBs in common with \citet{Guidorzi05}.  Some
discrepancies have been found.  Therefore, for reasons of homogeneity,
we have included in our sample the measured \tdurob\ and $V$ from
\citet{Guidorzi05} only for one event, GRB~000911.

% =======================================================
\section{A new tight correlation}
% =======================================================

Our aim is to find correlations among prompt $\gamma-$ray quantities
alone, with the smallest scatter.  For example, we would like to
explore the possibility to improve the \Liso--\Ep\ (\Eiso--\Ep)
``power'' (``energy'') correlations by introducing the prompt emission
variables \tdur, $V$, and $\Phi$, which partially characterize the
$\gamma$--ray light curve (see \S 4).  We remark that in our
statistical analysis we take into account the uncertainties of the
data.  Notice that the variables always refer to the GRB rest frame.
Our main criterion to judge about the goodness of the fit will be the
value of the reduced $\chi^2$.

% --------------------------------------------

\begin{figure}
\psfig{file=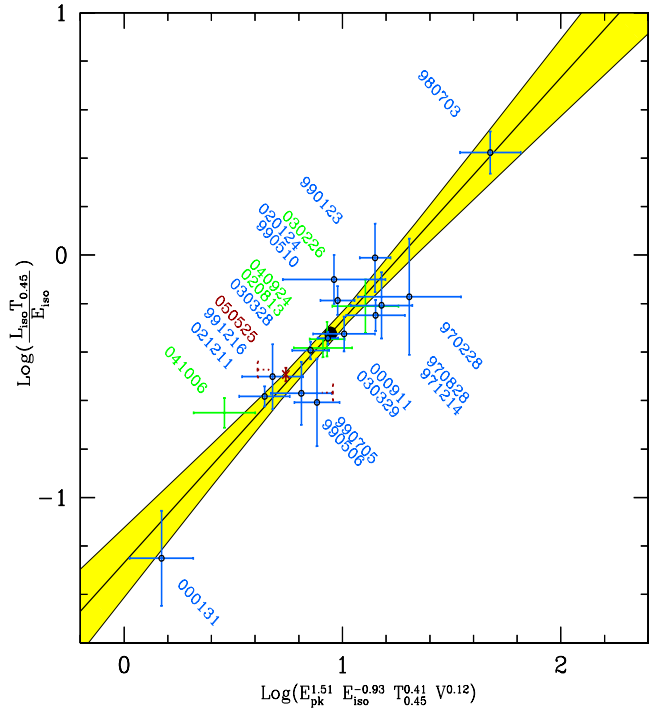,height=10cm,width=9cm}
\caption{Multi variable linear regression analysis based on the 15
GRBs of class 1 (open circles).  The best fit (solid line) corresponds
to Eq. \ref{Fi} and the shaded region is the 1$\sigma$ uncertainty of
this equation. The latter is computed in the barycenter of the data
points (solid black point). Color codes have the same meaning as in
Fig. \ref{fig1}.
\label{fig2}}
\end{figure}

Among other combinations, we look for a multi variable linear
correlation of Log$\Phi$ as a function of a combination of Log\Ep,
Log\Eiso, Log\tdur\, Log$V$ (\Liso\ is already contained in the
variable $\Phi$).  This approach also allows us to explore whether the
lightcurve variables, $\Phi$ and $V$, are correlated among them or
not.  Only the 15 high--quality (class 1) events with accurately
measured variability $V$ are used in the regression analysis.  The
best fit is:
\begin{eqnarray} 
{\Liso \tdur \over \Eiso} \, &=& \, 10^{-0.32\pm0.03}  
\left({\Ep \over 10^{2.38}\, {\rm keV}}\right)^{1.51\pm0.25} \times\nonumber \\
&~&\left({\Eiso \over 10^{2.95}\, {\rm erg}}\right)^{-0.93\pm0.16} \, 
\left({\tdur \over 10^{0.53}\, {\rm s}}\right)^{0.41\pm0.10}\times\nonumber \\
&~&\left({V \over 10^{-1.06}}\right)^{0.12\pm0.10},
\label{Fi} 
\end{eqnarray}
%EQUAZIONE VERIFICATA
with \chrs=0.62 (for 10 dof).  Fig. \ref{fig2} shows the data and the
best fit given by Eq. (\ref{Fi}) in a two dimensional diagram.  The
uncertainties in the data, showed with error bars, were taken into
account in the regression analysis.  The (yellow) shaded area
corresponds to the $1\sigma$ confidence interval of the regression
line (Eq.\ref{Fi}).  The data used for the fitting are marked with
open circles.  The class 2 events and GRB~050525 are also
plotted in Fig. \ref{fig2}, although they were not used in the fit.

% --------------------------------------------

\begin{figure*}
\psfig{file=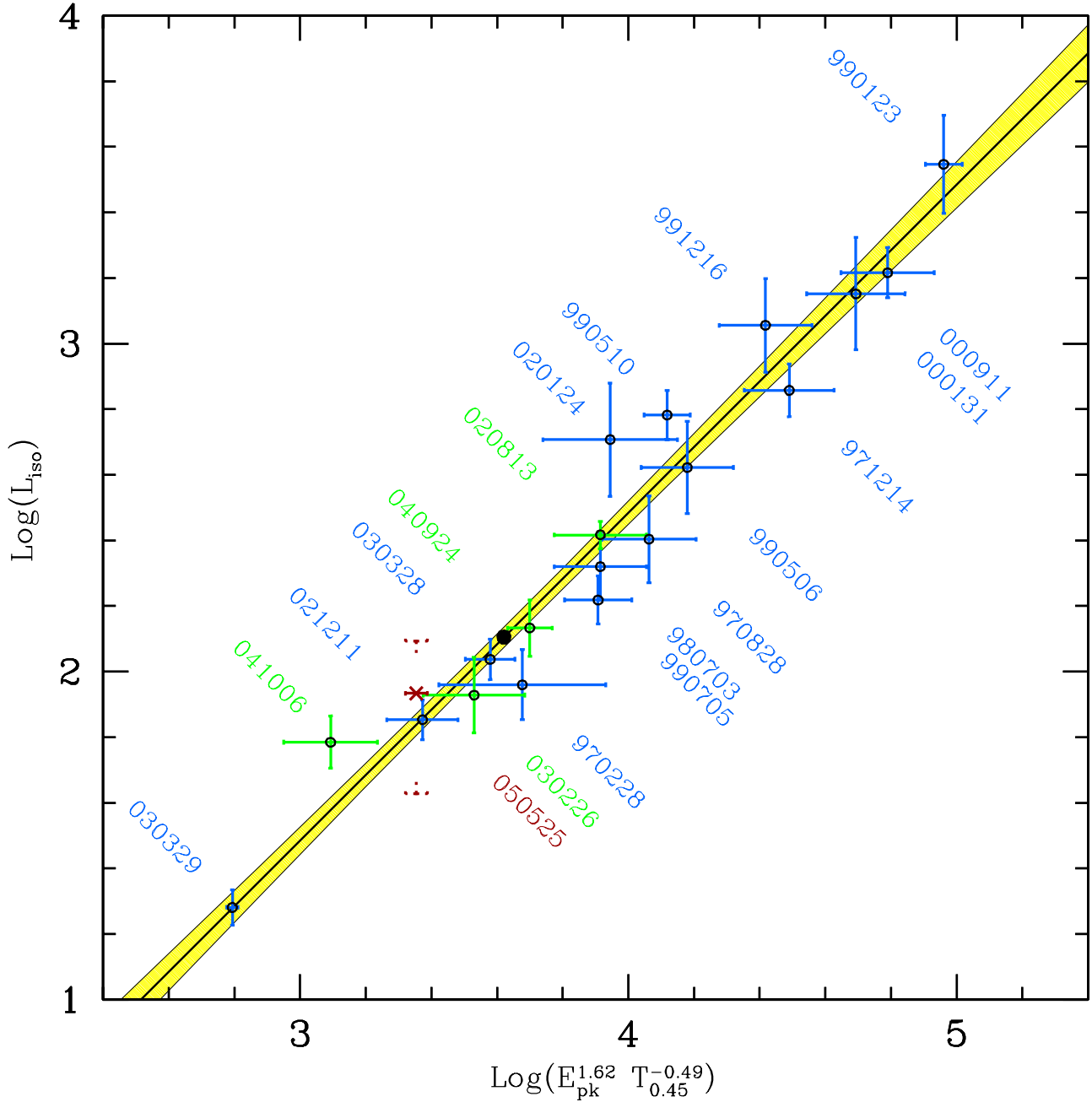,height=18.5cm,width=18.5cm}
\caption{ Multi variable linear regression analysis based on 19 GRBs
(class 1 [blue crosses] and class 2 [green crosses] events in
Tab. 3). The best fit, corresponding to Eq. \ref{Liso}, and its
1$\sigma$ uncertainty (computed in the barycenter of the data points -
solid black point) are represented by the solid line and by the shaded
(yellow) region, respectively.  using the same convention of
Fig. \ref{fig1}.
\label{fig3}}
\end{figure*}

From Eq. (\ref{Fi}) one sees that $\Phi$ is correlated with $V$ very
weakly, since its exponent is almost consistent with zero. We also
note that \Eiso\ is not an important variable in the correlation
represented by Eq. (\ref{Fi}), because it appears with almost the same
power on both sides of this equation.
%probably because it is redundant with \Liso\ and/or \Ep. 
Therefore, the significant variables are \Liso, \Ep\ and \tdur. 
In a first approximation, from Eq. (\ref{Fi}) we may infer that 
$\Liso \propto \Ep^{1.5} \tdur^{-0.5}$. 

We therefore proceed to perform the multi variable regression analysis
(taking into account the uncertainties) for Log\Liso\ as a function of
Log\Ep\ and Log\tdur.  For the same events used previously, we obtain
a fit with \chrs=0.67 (for 13 dof) that, compared with the \chrs=0.62
of the correlation given by Eq. (\ref{Fi}), implies again that the
variables $V$ and $\Eiso$ are redundant for the correlation in
Eq. \ref{Fi}.  In order to use as many data as possible, we also
performed the regression analysis including 4 events of class 2 (GRB
030429 lacks a measure of \tdurob), for a total of 19 events.
% None of these events is an outlier in the 
% \Eiso--\Ep--$t_{\rm break}$ correlation. 
The best fit is:
\begin{eqnarray} 
\Liso \, &=& \, 10^{52.11\pm0.03} 
\left({\Ep \over 10^{2.37}\, {\rm keV}}\right)^{1.62\pm0.08}
\times\nonumber \\
&~& \left({\tdur \over 10^{0.46}\, {\rm s}}\right)^{-0.49\pm0.07} 
\,\,\, {\rm erg\,\, s^{-1}}
\label{Liso} 
\end{eqnarray}
%EQUAZIONE VERIFICATA
with \chrs=0.70 (for 16 dof). 
Fig. \ref{fig3} shows the data, the best fit and its $1\sigma$ 
confidence interval (shaded area) in a two dimensional diagram. 
GRB~050525 was not used for the regression analysis but it is plotted in 
Fig. \ref{fig3}. 

% -----------------------------------------------

\begin{figure}
{\hskip -1 cm  \psfig{file=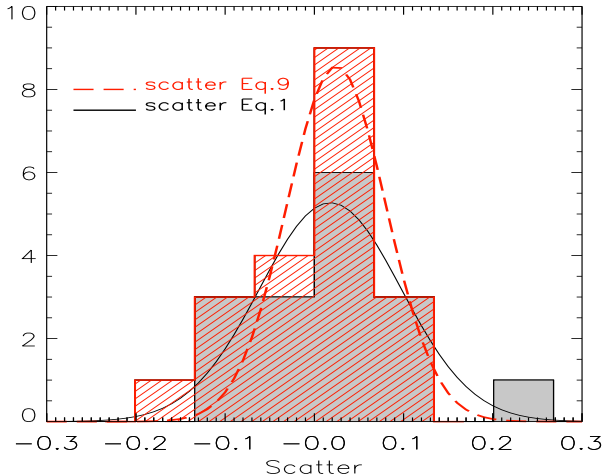,height=7cm,width=10cm}}
\caption{Histogram of the (logarithmic) scatter of the data points
around the best fit of Eq. \ref{Liso} (light gray) 
and Eq. \ref{Liang} (darker gray).
We superpose the two Gaussian fits.
\label{scatter}}
\end{figure}

The correlation \Liso--\Ep--\tdur\  is very tight, which can be appreciated 
from Fig. \ref{scatter}, where we show the scatter (taken as the orthogonal
distance of the data points to the best fit line).
Fitting the scatter distribution with a Gaussian yields
a (logarithmic) dispersion $\sigma=0.06$, smaller than the 
corresponding dispersion of points around the correlation
described by Eq. \ref{Liang} (a Gaussian yields $\sigma=0.08$).
% the small relative errors in the zero point and in the exponents.

The correlation of Eq. \ref{Liso}
is a significant improvement (i.e. it has a smaller scatter)
of the  $\Liso$--$\Ep$ correlation found by  
\citet[]{Yonetoku04}.  
In fact, for the updated version of the Yonetoku correlation, 
Ghirlanda et al. (2005) found a 1$\sigma$ dispersion of 0.25.
We have explored other multi variable correlations and the one given 
by Eq. (\ref{Liso}) was indeed the tightest with significant variables.

The fact that \chrs\ is smaller than 1 might be a consequence of
an overestimate in the observational uncertainties of the data as
well as of unaccounted correlation among the variable errors.  Indeed, 
we have assumed that the errors in the variables are uncorrelated (\S 2).  
If the errors instead have some degree of correlation among them (which is
very likely), then we are overestimating their contribution to \chrs.

Fig. \ref{fig3} shows the main result of this paper, namely the
existence of a very tight empirical relation among the peak luminosity
of the GRB (\Liso) and the prompt $\gamma$--ray quantities \Ep\ and
\tdur. This relation is as tight as the $\Eiso$--$\Ep$--$\tbreak$ one
(see \S 3).

\begin{figure*}
\psfig{file=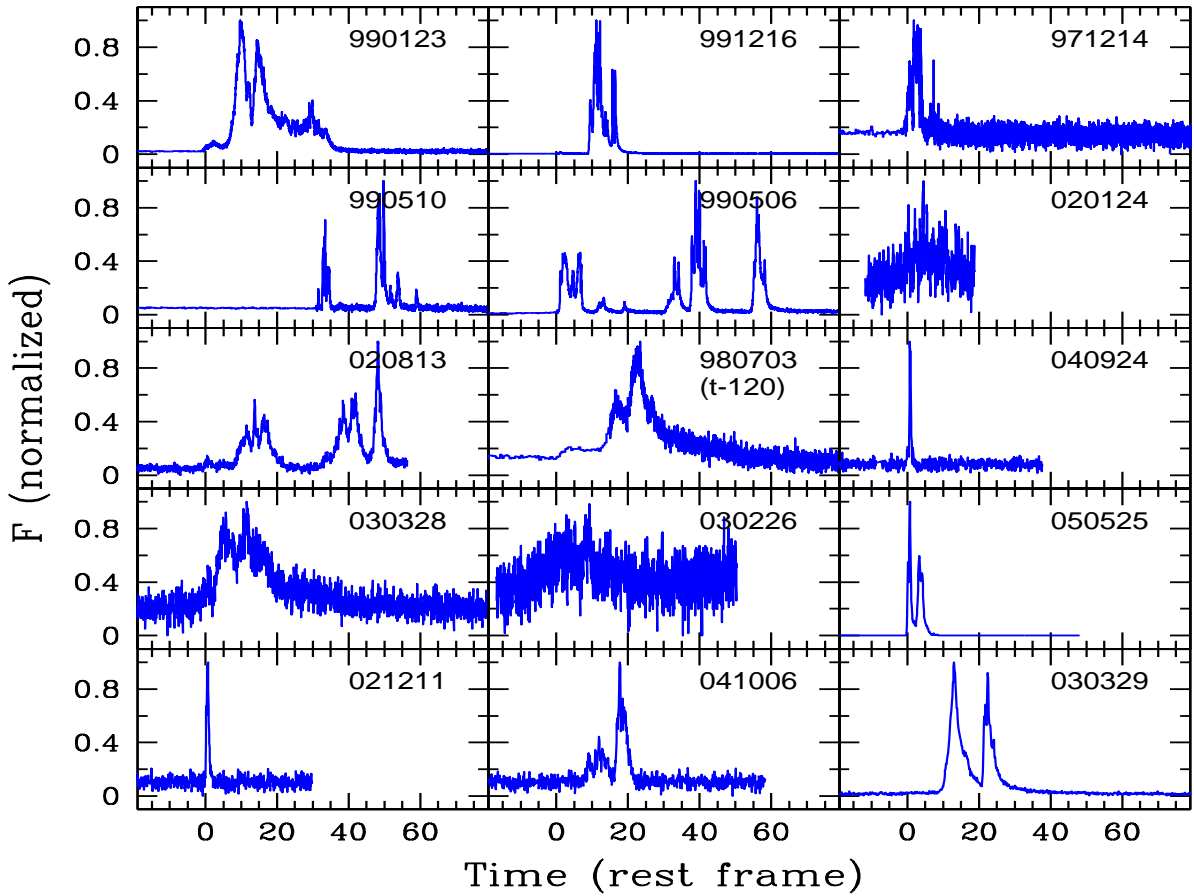,height=14cm,width=17cm}
\vskip -0.5 true cm
\caption{
The sequence of the available GRB light curves of our sample. The
order follows the distribution of GRBs along the sequence of 
Fig. \ref{fig3} from top--right to bottom--left. 
\label{fig4}}
\end{figure*}

In Fig. \ref{fig4} we display  the available GRB light--curves following
the same sequence of data points (from top--right to bottom--left) 
plotted in Fig. \ref{fig3}.  The different
patterns appear mixed, and there is no hint of any sequence.

% ---------------------------------------------------

We have also performed a linear regression analysis to find the best
fit in our sample for $\Liso$--$\Ep$--$T_{90}$, i.e. using the
variable $T_{90}$ at rest instead of \tdur. The obtained multivariable
correlation is similar to the one given by Eq. (\ref{Liso}) but with a
significant larger scatter.

\subsection{The lowest--luminosity GRBs in the \Liso--\Ep--\tdur\ diagram}

There are four GRBs that were not 
included in our sample: GRB~980425 (a rather peculiar burst),
GRB~990712 and GRB~010921 (they were not included in our 
sample because they have a too uncertain determination of $V$), 
and GRB~031203 (another peculiar burst, as GRB~980425).
All these bursts are low luminosity events.    
It is interesting to see a posteriori where they 
lie in the Log\Liso\ vs Log($\Ep^{1.62}\tdur^{-0.49}$) plot.

% -------------------------------------------------

\begin{figure}
\psfig{file=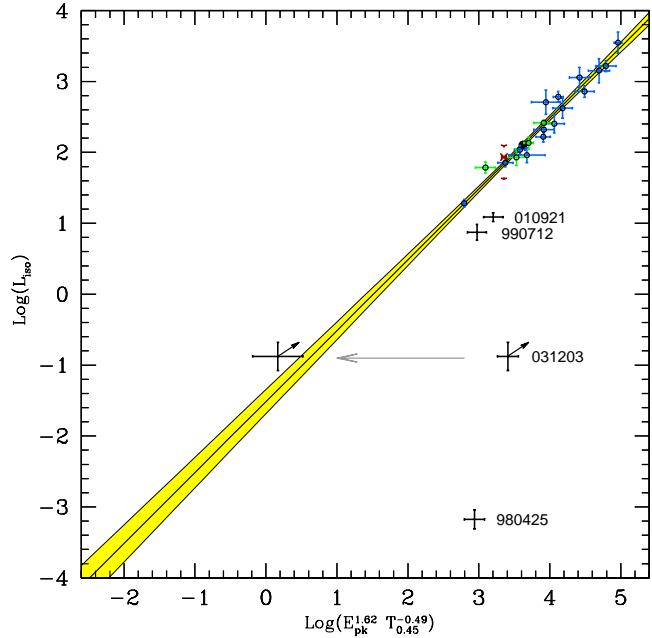,height=9cm,width=9cm}
\caption{
Similar to Fig. \ref{fig3} but including the  ``outliers'' 
GRB~980425, GRB~990712, GRB~010921 and GRB~031203. 
If the X--ray \Ep\ reported by \citet{watson06} is used, then the 
position of GRB~031203 (lower limit calculated for $ \Ep = 200 \pm 40 keV$) 
shifts close to the correlation given by Eq. \ref{Liso}.  
\label{fig5}}
\end{figure}

Fig. \ref{fig5} shows the same correlation of Fig. \ref{fig3} but we
include here these four low luminosity events.  Curiously enough, they
break the main correlation at its lowest end, below $\Liso\approx
10^{51.2}$ erg s$^{-1}$, giving rise to a kind of low luminosity
branch (determined mainly by a minimum value of \Ep\ and, at a less
extent, by a maximum value of \tdur).  
%It will be interesting to
%explore in more detail the potential physical implications of the
%information contained in Fig. \ref{fig5}.

In a recent paper, \citet{watson06} have shown that GRB~031203 might
have a soft X--ray spectral component, at $\sim 2$ keV.  This
component carries a fluence which is even higher than the
$\gamma$--ray one.  Therefore it is this soft component which can
determine the peak energy \Ep.

If we use \Ep\ of the X--ray component reported in \citet{watson06},
then the position of GRB~031203 in Fig. \ref{fig5} shifts to the left
as indicated by the arrow in Fig. \ref{fig5}.  In this case GRB~031203
lies close to the multi variable correlation given by
Eq. (\ref{Liso}).  Furthermore, the reported spectra of GRB~990712 and
GRB~010921 (\citet{amati02} and \citet{sakamoto05}, respectively),
seems peculiar: in the case of GRB~990712 the spectrum is rather flat
in $\nu F(\nu)$, making the determination of \Ep uncertain, and in the
case of GRB~010921 the spectrum has some local bumps at energies lower
than the reported \Ep\ (Barraud et al. 2003).  It is then possible
that also these two sources lie very close to the main correlation
shown in Fig. \ref{fig5}.

%=======================================================
\section{The \relL\ correlation as a redshift indicator}
%=======================================================

\begin{table}
\centering
\begin{tabular}{@{}lccc@{}}
%%\scriptsize
\hline\hline
GRB      & $z$    & $\bar{z}$&  $\Delta z/z$\\ 
\hline
 970228  & 0.695  & 0.940  &  0.353   \\
 970828  & 0.958  & 1.212  &  0.265   \\
 971214  & 3.420  & 4.801  &  0.404   \\
 980703  & 0.966  & 1.104  &  0.142   \\
 990123  & 1.600  & 1.358  & -0.151   \\
 990506  & 1.307  & 1.418  &  0.085   \\
 990510  & 1.619  & 1.210  & -0.253   \\
 990705  & 0.843  & 1.109  &  0.315   \\
 991216  & 1.020  & 0.823  & -0.193   \\
 000131  & 4.500  & 4.957  &  0.102   \\
 000911  & 1.058  & 1.173  &  0.108   \\
 020124  & 3.198  & 1.797  & -0.438   \\
 020813  & 1.255  & 1.235  & -0.016   \\
 021211  & 1.010  & 1.032  &  0.022   \\
 030226  & 1.986  & 2.398  &  0.208   \\
 030328  & 1.520  & 1.613  &  0.061   \\
 030329  & 0.169  & 0.170  &  0.008   \\
 040924  & 0.859  & 0.941  &  0.095   \\
 041006  & 0.716  & 0.547  & -0.236   \\
\hline\hline
\end{tabular}
\caption{ Spectroscopically measured redshifts ($z$) and
pseudo--redshifts ($\bar{z}$) derived according to the procedure
described in Sec.~7 for 20 GRBs of our sample.  $\Delta z/z$
represents the uncertainty of the derived pseudo--redshifts.}
\end{table}
% -------------------------------------------

The tightness of the multi variable \relL\ correlation (see
Fig. \ref{scatter}) encourages us to use it as new distance (redshift)
indicator for GRBs with no measured redshifts. In fact, the great
advantage of this correlation is that involves only prompt emission
quantities which can be easily derived from the $\gamma$--ray
lightcurve and spectra of GRBs with unknown $z$.
% and by assuming a cosmological model. 

% Making use of the measured prompt 
% $\gamma$--ray properties associated to the rest quantities entering 
% in the \relL\ relationship, pseudo--luminosities and pseudo--redshifts 
% for hundreds of GRBs could  be estimated.  
This way, we could have a statistically significant GRB
luminosity--redshift diagram to infer the
% (flux--limited) 
GRB luminosity function and the GRB formation rate history
\citep[e.g.,][]{Schaefer01, LFR02, Firmani04, Yonetoku04}.
% Here we estimate the accuracy of
% the pseudo-redshifts estimated on basis of the \relL\ relationship.

The bolometric corrected peak flux, in the observer frame, is
connected to the bolometric corrected isotropic luminosity \Liso\ by:
\begin{equation}
P=\frac{\Liso}{4\pi D_L^2(z)} 
\label{Pbol}
\end{equation}
where $D_L(z)$ is the luminosity distance.  The cosmological redshift
and time dilation effect impose that $\Ep = E_{\rm pk}^{\rm obs}(1+z)$
and $\tdur = \tdurob/(1+z)$. 
However, the use of \tdurob\ needs 
a particular care.
In fact, as mentioned in \S 2.1, \tdurob\
(and consequently $\rm T_{0.45}$)
is calculated from a light curve defined over an energy range which is
fixed in the rest frame of the GRB (see also R2001).  Using the
approximation given by R2001 that takes into account the narrowing of
the light curve temporal substructures at higher energies, we obtain
$\tilde{T}_{0.45}^{\rm obs}=\tdurob/(1+z)^{0.4}$, where
$\tilde{T}_{0.45}^{\rm obs}$ is measured 
% the 45\% effective time from the lightcurve 
in the same {\it energy range but defined now in the observer frame}. 
It follows that 
% the 45\% effective time 
\tdur\ measured in the {\it source rest frame} is $\tdur =
\tilde{T}_{0.45}^{\rm obs}/(1+z)^{0.6}$.

\begin{figure}
\psfig{file=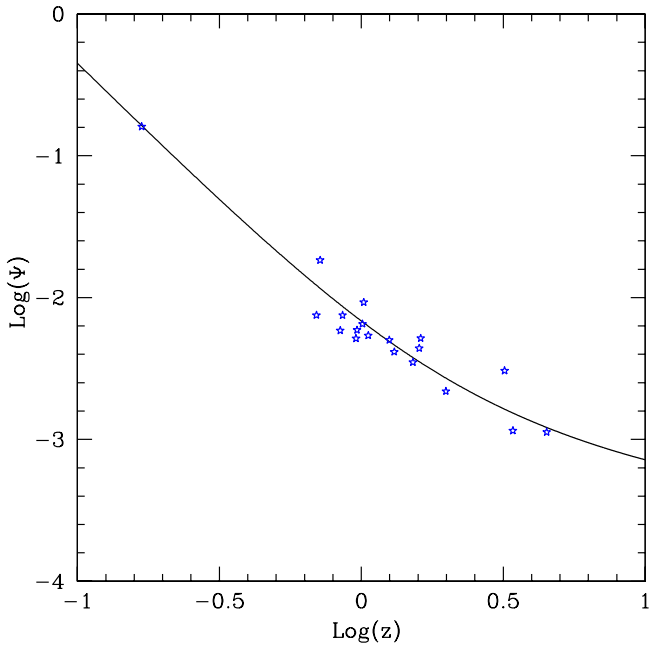,height=9cm,width=9cm}
\caption{
The line shows the pseudoredshift given by Eq. \ref{pseudoz1}, while the 
data show the observed redshifts for the 19 GRBs of the sample of 
Fig. \ref{fig3}.  
\label{fig6}}
\end{figure}

Once we know how the rest frame quantities $L_{\rm iso}$, $E_{\rm pk}$
and \tdur\ transform with the redshift $z$, we may express the \relL\
relation (Eq. \ref{Liso}) in terms of the observer frame equivalent
quantities:
\begin{eqnarray} 
P 4\pi D^2_L(z) &~& \, =\,  10^{52.11\pm0.03} 
\left[{E_{\rm pk}^{\rm obs}(1+z) \over 10^{2.37}\, 
{\rm keV}}\right]^{1.62\pm0.08}
\times\nonumber \\
&~& \left[{ \tilde{T}_{0.45}^{\rm obs}\over (1+z)^{0.6}\, 10^{0.46}\, 
{\rm s}}\right]^{-0.49\pm0.07} 
\,\,\, {\rm erg\,\, s^{-1}}
\label{Liso_obs} 
\end{eqnarray}
We indicate the ratio between observer frame measured quantities as
$\Psi$ and we isolate the redshift dependences in the term $f(z)$:
\begin{eqnarray}
f(z)=\frac{(1+z)^{1.91}}{4\pi D_L^2} \\
\Psi \equiv \frac{ P (\tilde{T}_{0.45}^{\rm obs})^{0.49} }
{{E_{\rm pk}^{\rm obs}}^{1.62}},
\end{eqnarray}
%
% elated to those entering in the \relL\ relationship
% (Eq. \ref{Liso}), we may rewrite this relation in terms of the
% observable quantities and of a function that depends on $z$:
%
so that Eq.~\ref{Liso_obs} reduces to:
\begin{equation}
f(z) = C \Psi,
\label{pseudoz}
\end{equation}
where $C$ is fixed to $10^{6.48}$ such that $D_L$, $P$, $E_{\rm pk}^{\rm obs}$, 
and $\tilde{T}_{0.45}^{\rm obs}$ are
expressed in unities of Gpc, erg cm$^{-2}$ s$^{-1}$, keV and sec, respectively.

% $\Psi$ is obtained from directly observed quantities.
% We did not take into account error propagation in Eq. (\ref{pseudoz})

For the \LCDM\ cosmological model assumed in this paper, the inversion
of Eq. (\ref{pseudoz}) allows us to estimate the pseudo--redshift
$\bar{z}$ of a given GRB whose $P$, $E_{\rm pk}^{\rm obs}$, and
$\tilde{T}_{0.45}^{\rm obs}$ have been measured from the prompt
emission $\gamma$--ray light curve solely:
\begin{equation}
\bar{z} = f^{-1}(C \Psi).
\label{pseudoz1}
\end{equation}

\begin{figure}
\vskip -0.5 true cm
\psfig{file=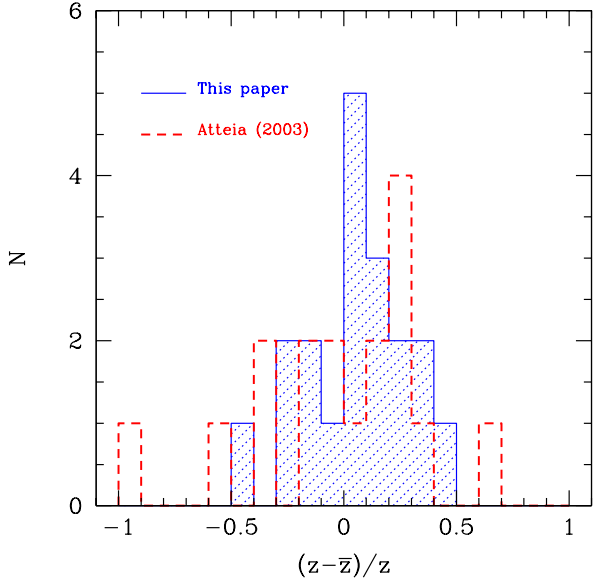,height=9cm,width=9cm}
\vskip -0.5 true cm
\caption{
Histograms of the values of $\Delta z/z$ obtained with
the method discussed in this paper  (hatched, 19 GRBs) and by
by Atteia (2003, dashed line, 17 GRBs).
\label{atteia}}
\end{figure}

In Fig. \ref{fig6} we plot Eq. (\ref{pseudoz}) as Log$\Psi$
vs. Log($z$) (solid line).  
The sample of GRBs used to derive the
\relL\ correlation are also plotted (stars).  
The notably small scatter of the data points around the curve is 
a proof of the reliability of the redshift indicator inferred from 
the \relL\ correlation.  
In Table 5 we give for the 19 GRBs of our sample the
spectroscopically measured $z$, the pseudo--redshifts $\bar{z}$
derived from Eq. \ref{pseudoz1}, and the relative uncertainty between
these two quantities.  
We note that the 68\% of our pseudo redshifts (i.e. 13 out of 19 sources)
are within the 25\% of the real value. 
% on average the uncertainty
% associated to the pseudo--redshifts is of only 16\%, with a
% standard deviation of $\pm 13\%$, 
In Fig. \ref{atteia} we compare the distribution of the $\Delta z/z$ values
obtained with Eq. \ref{pseudoz1} with the one found by Atteia (2003).
One can see that the our distribution is somewhat narrower, and especially 
that the central value is closer to zero.

% This is considerably lower than what found with other proposed redshift 
% indicators (Atteia 2003).

Moreover, as can be inferred from the data reported in
Table 4, the relative uncertainty on the derived pseudo--redshifts
does not seem to be correlated with the redshifts. This suggests that
the \relL\ correlation can be used to estimate either low and high
values of $z$.

Finally we note that the \tdurob\ parameter can be derived from the
prompt emission lightcurve straightforwardly. On the other hand, the use
of Eq.~\ref{pseudoz1} to estimate $z$ requires the measure of the GRB peak
spectral energy $E_{\rm pk}^{\rm obs}$ from the fit of a broad--band
spectrum.
% The average relative error is 16\% with a standard deviation of $\pm
% 13\%$.  This is a significant improvement on previous GRB redshift
% indicators.

%=======================================================
\section{Theoretical implications and challenges}
%=======================================================

\subsection{Comparison with other correlations}

The most important correlation found in this paper among $L_{\rm iso}$,
\Ep\ and $T_{0.45}$ can be compared with the other tight correlations
between the emitted energy or luminosity, \Ep\ and \tbreak.
Consider first the Amati relation between $E_{\rm iso}$ and \Ep.
It has a relatively large scatter, which ``collapses" when calculating
the collimation corrected emitted energy \Ecol.
In this case, the ``collapsing parameter" going from the Amati to the 
Ghirlanda relation is the jet opening angle, measured by \tbreak.
If indeed $\theta_j$ is well measured by \tbreak, then the existence
of the Ghirlanda relation implies the existence of the phenomenological 
\Eiso--\Ep--\tbreak\ tight relation (Eq. \ref{Liang}) (and vice--versa).

Consider now the Yonetoku relation between $L_{\rm iso}$ and \Ep.
As detailed in Ghirlanda et al. (2005), it has a scatter which is 
comparable to the scatter of the Amati relation.
But if we now calculate $L_\gamma$ (correcting for the jet opening angle),
then we {\it do not} find a correlation as tight as the Ghirlanda one.
Therefore the ``collapsing parameter" which worked for the total energetics
does not work for the peak luminosity.
The correlation found in this paper demonstrates that a 
better ``collapsing parameter" for the peak luminosity is $T_{0.45}$.

From this we can already conclude that $T_{0.45}$ {\it is not}
equivalent to the jet opening angle: if it were, then both $T_{0.45}$ 
and $\theta_j$ (or \tbreak) should work as collapsing parameters without
``caring" if dealing with total prompt energetics or peak luminosities.

Therefore the two tight empirical multi variable correlations 
\Liso--\Ep--\tdur\ (Eq. \ref{Liso}) and  \Eiso--\Ep--\tbreak\ 
(Eq. \ref{Liang}) represent two different aspects of the GRB physics.

\subsection{``Local" versus ``global" properties}

It is of some interest to make a distinction between {\it local} and
{\it global} properties of GRBs.
If the fireball has a semi--aperture angle $\theta_j>1/\Gamma$,
we can see only a portion of the emitting surface (or volume).
The peak flux, the fluence and the timescale \tdur\ are defined
as local quantities since they correspond only to the ``observable"
surface (or volume). 
This implies that also \Liso\ and \Eiso\ are basically local quantities.
We can then introduce the concept of ``local brightness"
characterizing the flux emitted by the visible fraction of the fireball.
On the other hand, the jet opening angle $\theta_j$ (and consequently \tbreak)
is a global quantity.
This implies that $E_\gamma$ and the Ghirlanda relation
are expressions of a global property of GRBs.
On the other hand, the \Liso--\Ep--\tdur\ relation involves
the local brightness (entering in the definition of \Liso),
which is a local property.
Furthermore, \Liso\ is an instantaneous quantity, not time integrated
(contrary to \Eiso\ or $E_\gamma$). 

% Under the assumption that there is
% a direct connection between \tbreak\ and \tj\ (see \S 3),
% this implies that \tbreak\ is a global quantity, and consequently
% also $E_\gamma$ and $L_{\gamma}$.
% As a consequence, the \Eiso--\Ep--\tbreak\ and the Ghirlanda 
% correlations between \Ecol\ and \Ep\ link global and local quantities.
% On the other hand, the \Liso--\Ep--\tdur\ relation 
% links only local properties. 

We now discuss some implications of the combination of the
\relE\ and the \relL\ relations. 
% We therefore derive the conditions which makes these
% two relations mutually consistent.
We can divide both sides of Eq.~\ref{Liso} by 
$\Phi$ (see Eq.~\ref{Fi}) so that \Eiso\ becomes the
dependent variable in Eq.~\ref{Liso} as it is in Eq.~\ref{Liang}.
A more rigorous result can be obtained by fitting (with the multi
variable regression technique) Log\Eiso\ as a function of Log\Ep\,
Log\tdur\ and Log$\Phi$. The best fit, obtained for our sample
(including class 1 and 2 GRBs for a total of 19 events), is:
\begin{eqnarray}
\Eiso &=& 10^{52.90\pm0.03} 
\left({\Ep \over 10^{2.36}\, {\rm keV}}\right)^{1.63\pm0.08}\times 
\nonumber \\
&~&
\left({\tdur \over 10^{0.46}\, {\rm s}}\right)^{0.51\pm0.07}
\left({\Phi \over 10^{-0.34}}\right)^{-1.00\pm0.15}\,\, {\rm erg}
\label{LisoFi}
\end{eqnarray}
% EQUAZIONE VERIFICATA 
with \chrs = 0.64 (for 15 dof).  
This multi variable correlation is completely equivalent to that 
of Eq. (\ref{Liso}).
% and it can be directly compared to the correlation given 
% by Eq. (\ref{Liang}).  In fact, 
By eliminating \Ep\ in Eqs. (\ref{LisoFi}) and (\ref{Liang}), we
can express
% \footnote{
% We do not propagate the errors on the correlations´ parameters 
% because of the complexity and uncertainty in the error 
% mutual--correlations.}  
\tbreak\ as a function of of
$\Phi$ and \tdur\ (the dependence on \Eiso\ becomes negligible):
\begin{equation}
\tbreak \approx 4  \left({\Phi^2 \over \tdur } \right)^{0.6} \,\, 
{\rm days}
\label{connect}
\end{equation}
%EQUAZIONE VERIFICATA
Eq.~\ref{connect} establishes a clear connection between prompt
emission local quantities ($\Phi$ and \tdur),  which are related to
the $\gamma$--ray lightcurve variability, and an afterglow emission
global quantity, i.e. \tbreak.

LZ2005 suggested that \tbreak\ could be related to the prompt
($\gamma$--ray) emission rather than to the afterglow one.  \tbreak\
enters in the Ghirlanda correlation by means of the jet opening
angle (see \S 3), and, as discussed in LZ2005, it raises the question
of how a global quantity (i.e. $\theta_{\rm j}$) conspires with \Eiso\
to affect \Ep, which, instead, are both local quantities. 

It could be that \tbreak\ is determined by only prompt emission local
quantities. This is as also suggested by the reported (\citet{SG02})
correlation between the pulse lag of the prompt $\gamma$--ray
emission and \tbreak. This might justify the existence of an empirical
correlation between the jet angle (derived from \tbreak) and the local
properties \Eiso\ and \Ep. Therefore, our finding that \tbreak\ is
closely related to the prompt lightcurve quantities $\Phi$ and \tdur\
(Eq.~\ref{connect}) could be the basis of the Ghirlanda and the
\relE\ correlations.

%[Under the assumption of the uniform jet model, the correlation \relE\
%is less scattered than the Ghirlanda one  because it does not 
%include a dependence on uncertain quantities ($n$ and $\eta_\gamma$ for the HM case, 
%and $A$ and $\eta_\gamma$ for the WM case, see \S 3), which probably vary from burst to 
%burst, but that are assumed constant in the Ghirlanda relations. In this case, 
%the relationship Eq. (\ref{connect}) could serve
%as an independent way to estimate the unknown parameters   
%$n$ or $A$ and $\eta_\gamma$ related to the circumburst medium.  (...)]

\subsection{The \relL\ correlation in the comoving frame}

Another important aspect concerns the transformation
of the relation \Liso--\Ep--\tdur\ from 
the GRB rest frame to the fireball comoving frame.
Due to blueshift, \Ep\ and the comoving (primed) $E_{\rm pk}^\prime$
are related by $\Ep=E_{\rm pk}^\prime\delta$, where 
$\delta=1/[\Gamma(1-\beta\cos\xi)]$ is the relativistic Doppler factor,
and $\xi$ is the angle between the line of sight and the
velocity vector. 
Similarly, we have $T_{0.45}=T^\prime_{0.45}/\delta$.
For the isotropic equivalent luminosity we have two possible
transformations, according if the fireball is collimated in an angle
$\theta_j>1/\Gamma$ (``standard" fireball) or if its relevant aperture
angle $\theta_j<1/\Gamma$ [e.g.it is made by 
``bullets" (Heinz \& Begelman 1999;
sub--jets (Toma, Yamazaki \& Nakamura, 2005); 
or cannonballs  (e.g. Dar \& De R\`ujula 2004)].
In the former case we have $L_{\rm iso} = \delta^2 L^\prime_{\rm iso}$,
while in the latter $L_{\rm iso} = \delta^4 L^\prime_{\rm iso}$.
This is due to the fact that, for $\theta_j>1/\Gamma$, radiation is collimated
in a cone of semiaperture angle $\theta_j$, while, if $\theta_j<1/\Gamma$,
the collimation angle becomes $1/\Gamma$.

For ease of discussion, let us approximate  Eq. \ref{Liso} as
$ L_{\rm iso} \propto  E_{\rm pk}^{3/2} T_{0.45}^{-1/2}$.
It is clear that, for $\theta_j>1/\Gamma$, the same relation holds
in the comoving frame: {\it the $\delta$ factors cancel out.}
Instead, for $\theta_j<1/\Gamma$, we have
$ L^\prime_{\rm iso} 
\propto \delta^{-2}{(E^\prime_{\rm pk})}^{3/2} {(T^\prime_{0.45})}^{-1/2}$:
there is a rather strong dependence on $\delta$.

In other words, Eq. \ref{Liso} is ``Lorentz invariant" 
only for ``normal" fireballs (i.e. those with $\theta_j>1/\Gamma$).
In the opposite case ($\theta_j<1/\Gamma$) if the $\Gamma$ factor varies
more than a given (small) factor from event to event, then
we have a conflict with the tightness of our relation. 
We can therefore conclude that the standard fireball scenario
is favored with respect to models in which we see the entire
emitting surface (cannonballs, sub--jets and bullets, of
angular extension $\theta<1/\Gamma$).

If the ``Lorentz invariance" can be used as a discriminating guide
among the two possible forms of the Ghirlanda relation (favoring its
``wind--like" version, which linearly links $E_\gamma$ and \Ep),
then we have, from that relation, that the number of relevant 
photons of the prompt emission (those contributing the most to $E_\gamma$) 
must be the same in different bursts (see Eq. 6 and Nava et al. 2006). From 
the \relL\ relation found in this paper, instead, we derive 
a relation between the local (and instantaneous) maximum brightness of
the fireball, the \tdur\ timescale and \Ep. 
This relation, in the comoving frame, reads
\begin{equation}
r^2 F^\prime \propto 
{ ({E^\prime_{\rm pk}})^{3/2} \over {(T^\prime_{0.45})}^{1/2} } \, \to \,
r^2 \dot n^\prime_\gamma \sim r^2 {F^\prime\over E^\prime_{\rm pk}} \propto 
\left[{ E^\prime_{\rm pk} \over T^\prime_{0.45}}\right]^{1/2} 
\label{com}
\end{equation}
where $r$ is the radius of the fireball when it emits the peak brightness
$F^\prime$ and the photon peak flux $\dot n^\prime_\gamma$. From this perspective, 
we can see that the Ghirlanda relation
and the relation found here (Eq. 9) are complementary, describing two
different aspects (albeit hopefully related) of the GRB physics.

%=======================================================
\section{Conclusions}
%=======================================================

GRBs are extremely interesting objects: (i) their nature and physics
are still a mystery to solve, and, (ii) GRBs, being the most powerful
explosions in the Universe, can serve as a valuable cosmological tool.

An important step forward in the study of GRBs has been recently achieved
through the discovery of several correlations among their
intrinsic properties.  

Through the analysis of a sample of 22 GRBs, we have discovered a new
very tight correlation among {\it $\gamma$--ray prompt emission
properties alone} (Sec.~5).  The rest frame prompt--emission
quantities that we have correlated are: the bolometric corrected
energy and luminosity \Eiso\ and \Liso, the prompt emission spectrum
peak energy \Ep, the variability of the $\gamma$--ray lightcurve $V$,
and the ``high signal'' timescale \tdur (as defined in Sec.~4.1). We have
adopted in this paper an improved method to estimate $V$ and \tdur\
from the prompt emission lightcurve. Moreover, we adopted an integral
measure of the light curve variability as defined through the ratio
$\Phi=\Liso \tdur/\Eiso$.
The number of objects available to find these correlation 
is unfortunately still small, and a 
robust physical explanation is still to be found.
Bearing that in mind, the main conclusions of our study are:
\begin{itemize}
\item 
Long GRBs with measured redshifts obey a very tight rest frame
multi variable correlation which involves three prompt emission
quantities: $\Liso\propto E_{\rm pk}^{1.62} T_{0.45}^{-0.49}$
(Eq. [\ref{Liso}] and Fig. 3). 
%The best fit results in \chrs = 0.7 and
%for 16 dof (\chrs\ is less than 1 probably because of an overestimate
%of the uncertainties in the error propagation). 
The scatter of this correlation is comparable to that of the \Eiso,
\Ep\ and \tbreak\ (LZ2005; Fig. 1). Nonetheless, the newly found
\relL\ correlation {\it involves only prompt--emission quantities}.
\item 
The four GRBs (980425, 990712, 010921 and 031203) which populate
  the low--luminosity tail of the $L_{\rm iso}$ distribution of the
  bursts of our sample, were not included in the derivation of the
  above correlation. This is because of their arguable nature and/or high
  uncertainties in their observational parameters. However, with
  respect to the \relL\ correlation, they define a low luminosity
  ``branch'' (Fig. 6). Whether the origin of this branch is physical
  or merely a consequence of the lack of spectral data at low energies
  (as for GRB 031203), is an intriguing question which requires
  further analysis.
\item 
Other tight correlations among a larger number of prompt
emission quantities were also found by our multi variable analysis.
For example, we reported a correlation of $\Phi$ as a function of \Ep,
\Eiso, \tdur\ and $V$ (\chrs = 0.62; Eq.~\ref{Fi} and Fig. 2), and of
\Eiso\ as a function of \Ep, \tdur\ and $\Phi$ (\chrs = 0.63;
Eq.~\ref{LisoFi}). However, these extra variables (with respect to the
\relL\ correlation), only slightly improve the best fit
\chrs, at the cost to include a larger number of variables.
%introduce a larger scatter with respect to the \relL\ correlation. 
% This correlation presents also a (very weak) dependence of the 
% variability $V$.
% in the edge--on projection of a potential fundamental plane of prompt
% GRB properties.
\item 
The tightness of the multi variable correlation \relL\ allows us to use
it as a cosmology--dependent redshift indicator for GRBs. 
For the concordance cosmological model used here, we presented the formulas
necessary to infer pseudo--redshifts given three observables: 
the bolometric corrected $P$, $E_{\rm pk}^{\rm obs}$, and \tdurob. 
This method represents an improvement with respect to previous 
methods.
% The average
% relative error of our pseudo-redshifts is of 16\% with a standard
% deviation of $\pm 13\%$, which significantly improves 
\item 
We have re--derived the \relE\ relation (LZ2005) for the 15 GRBs out 
of our sample with known \tbreak. 
Then, we connected this relation to one equivalent
to the \relL\ relationship (\Eiso[\Ep,\tbreak,$\Phi$]) and found
$\tbreak\approx 4(\Phi^2 / \tdur)^{0.6}$ days.
This dependence establishes a connection between prompt local
quantities ($\Phi$ and \tdur, related to the $\gamma$--ray lightcurve
variability), and a global quantity associated to the afterglow 
(\tbreak), suggesting that \tbreak\ (and \tj) could be determined 
mainly by the prompt emission local properties.
\item 
In the context of the standard fireball scenario, the \relL\
correlation is roughly ``Lorentz invariant". This allows us to go
deeper and relate it to the local comoving fireball surface brightness
associated to the fireball area that the observer sees.  
Therefore, while the Ghirlanda relations seems to be related to a global
feature of the fireball (the number of emitted photons), the \relL\
relationship could be describing a local feature of the fireball
related to its brightness.  
% In the context of the cannonball scenario, the \relL\ relationship is not
% Lorentz invariant and its tightness might pose a difficulty for this
% scenario (?).
\end{itemize}

The results found in this paper reveal the existence of tight
correlations among quantities related to the prompt emission of
long GRBs. On one hand, these correlations will contribute to understand 
better the physics of GRBs. On the other hand, they can be used
to infer pseudo--redshifts with good accuracy for hundreds of GRBs
or, for those events with known $z$, serve as ``standard candles''
for cosmographic purposes. 
The main challenge is the acquirement
of observational data related to the $\gamma$--ray prompt light curves
and spectral distributions. In particular, detector flux sensitivities
extending to energies as high as 500--1000 keV are crucial,
also for revealing the effects of possible observational bias.

\section*{Acknowledgments}
%\acknowledgments

We thank the anonimous referee for insightful comments and
Davide Lazzati for discussions.
We also thank Giuseppe Malaspina for technical support and Yair Krongold
for useful advice on X--ray spectrum analysis. V.A-R. gratefully
acknowledges the hospitality extended by Osservatorio Astronomico di
Brera. We thank the italian INAF and MIUR for funding (Cofin 
grant 2003020775\_002).

%%%%%%%%%%%%%%%%%%%%%%%%%%%%%%%%%%%%%%%%%%%%%%%%%%%%%%%%%%%%%%%%%%%%%%

\end{document}